# Automating Legal Research through Data Mining

M.F.M Firdhous,

Faculty of Information Technology,
University of Moratuwa,
Moratuwa,
Sri Lanka.
Mohamed.Firdhous@uom.lk

*Abstract* —**The term legal research generally refers to the process of identifying and retrieving appropriate information necessary to support legal decision-making from past case records. At present, the process is mostly manual, but some traditional technologies such as keyword searching are commonly used to speed the process up. But a keyword search is not a comprehensive search to cater to the requirements of legal research as the search result includes too many false hits in terms of irrelevant case records. Hence the present generic tools cannot be used to automate legal research.**
**This paper presents a framework which was developed by combining several 'Text Mining' techniques to automate the process overcoming the difficulties in the existing methods. Further, the research also identifies the possible enhancements that could be done to enhance the effectiveness of the framework.**

## I. INTRODUCTION

Legal research is the process of identifying and retrieving information necessary to support legal decision-making. In its broadest sense, legal research includes each step of a course of action that begins with an analysis of the facts of a problem and concludes with the application and communication of the results of the investigation [1].

The processes of legal research vary according to the country and the legal system involved. However, legal research generally involves tasks such as finding primary sources of law, or primary authority, in a given jurisdiction (cases, statutes, regulations, etc.), searching secondary authority for background information about a legal topic (law review, legal treatise, legal encyclopedias, etc.), and searching non-legal sources for investigative or supporting information. Legal research is performed by anyone with a need for legal information including lawyers, law librarians, law students, legal researchers etc., Sources of legal information range from decided cases, printed books, to free legal research websites and information portals [2].

Manually performing legal research is time consuming and difficult. Because of that, some traditional tools have been introduced. There are essentially only two types of tools which help users find legal materials in the Internet, they are commonly known as catalogs and search engines. Combinations of catalogs and search engines in the same site are now becoming more common, and such combinations are often referred to as 'portals'. Despite the existence of these research aids, finding legal information on the internet is surprisingly difficult, partly because neither catalogs nor search engines used alone can provide a satisfactory solution. A general keyword search contains too much false hits. It is difficult to make searches precise enough to find only the relevant information.

In this case, the search should not be just a keyword search; instead an intelligent search should be carried out according to the meaning of the search text. This research presents a methodological framework based on text mining to automate legal research by focusing on the retrieval of exact information specifically necessary for legal information processing. The proposed approach uses a term-based text mining system and a vector space model for the development of the framework.

## II. RELATED WORK

Data mining is the process of sorting through large amounts of data and picking out relevant information. It is usually used by business intelligence organizations, and financial analysts, but is increasingly being used in the sciences to extract information from the enormous data sets generated by modern experimental and observational methods. It has been described as "the nontrivial extraction of implicit, previously unknown, and potentially useful information from data" and "the science of extracting useful information from large data sets or databases" [3].

Text mining, sometimes alternately referred to as text data mining, refers generally to the process of deriving high quality information from text. High quality information is typically derived through the division of patterns and trends through means such as statistical pattern learning. Text mining usually involves the process of structuring the input text (usually parsing, along with the addition of some derived linguistic features and the removal of others, and subsequent insertion into a database), deriving patterns within the structured data, and finally evaluation and interpretation of the output. 'High quality' in text mining usually refers to some combination of relevance, novelty, and interestingness. Typical text mining tasks include text categorization, text clustering, and concept/entity extraction, production of granular taxonomies, sentiment analysis, document summarization, and entity relation modeling [4–5].

### A. Dependency Analysis based Text Mining

One of the approaches in text mining is dependency based analysis. In [6–8] several methodologies in this area are presented. In [6], a text simplification approach is presented, whereas [7–8] present dependency analysis. Long and complicated sentences pose various problems to many state-of-the-art natural language technologies. For example, in parsing, as sentences become syntactically more complex, the number of parses increases, and there is a greater likelihood for an incorrect parse. In machine translation, complex sentences lead to increased ambiguity and potentially unsatisfactory translations. Complicated sentences can also lead to confusion in assembly/use/maintenance manuals for complex equipment [6].

Articulation-points are defined to be those points where sentences may be split for simplification. Segments of a sentence between two articulation points may be extracted as simplified sentences. The nature of the segments delineated by the articulation points depends on the type of the structural analysis performed. If the sentences are viewed as linear strings of words, articulation points can be defined to be, say, punctuation marks. If the words in the input are also tagged with part of speech information, sentences can be split based on the category information, for instance at relative pronouns, with part of speech information, subordinating and coordinating conjunctions may also be detected and used as articulation points. However, with just this information, the span of the subordinating/coordinating clause would be difficult to determine. On the other hand, if the sentence is annotated with phrasal bracketing, the beginnings and ends of phrases could also be articulation points.

The sentences in the training data are first processed to identify phrases that denote names of people, names of places or designations. These phrases are converted effectively to single lexical items. Each training sentence $S_i$, along with its associated j (simplified) sentences $S_{i1}$ to $S_{ij}$, is then processed using the Lightweight Dependency Analyzer (LDA) [7].

The resulting dependency representations of $S_i$ and $S_{i1}$ through $S_{ij}$ are 'chunked'. Chunking collapses certain substructures of the dependency representation (noun phrases and verb groups) and allows defining the syntax of a sentence at a coarser granularity. Chunking also makes the phrasal structure explicit, while maintaining dependency information. Thus, this approach has the benefit of both phrasal and dependency representations.

LDA is a heuristic based, linear time, deterministic algorithm which is not forced to produce dependency linkages spanning the entire sentence. LDA can produce a number of partial linkages since it is driven primarily by the need to satisfy local constraints without being forced to construct a single dependency linkage that spans the entire input. This, in fact, contributes to the robustness of LDA and promises to be a useful tool for parsing sentence fragments that are rampant in speech utterances exemplified by the switchboard corpus [7].

### B. Text Mining at the Term Level

Most efforts in Knowledge Discovery in Databases (KDD) have focused on knowledge discovery in structured databases, despite the tremendous amount of online information that appears only in collections of unstructured text. At abstract level, KDD is concerned with the methods and techniques for making sense of data. The main problem addressed by the KDD process is mapping low-level data into other forms that might be more compact, more abstract, or more useful. At the core of the process is the application of specific data-mining methods for pattern discovery and extraction. Previous approaches to text mining have used either tags attached to documents [9] or words contained in the documents [10].

The exploitation of untagged, full text documents therefore requires some additional linguistic pre-processing, allowing the automated extraction from the documents of linguistic elements more complex than simple words. Normalized terms are used here, i.e. sequences of one or more lemmatized word forms (or lemmas) associated with their part-of-speech tags. "stock/N market/N" or "annual/Adj interest/N rate/N" are typical examples of such normalized terms [11]. In [11], an approach is presented to text mining, which is based on extracting meaningful terms from documents. The system described in this paper begins with collections of raw documents, without any labels or tags. Documents are first labeled with terms extracted directly from the documents. Next, the terms and additional higher-level entities (that are organized in a hierarchical taxonomy) are used to support a range of KDD operations on the documents. The frequency of co-occurrence of terms can provide the foundation for a wide range of KDD operations on collections of textual documents, such as finding sets of documents whose term distributions differ significantly from that of the full collection, other related collections, or collections from other points in time.

The next step is the Linguistic Preprocessing that includes Tokenization, Part-of-Speech tagging and Lemmatization. The objective of the Part-of-Speech tagging is to automatically associate morpho-syntactic categories such as noun, verb, adjective, etc., to the words in the document. In [12], a Transformation-Based Error-Driven Learning approach is presented for Part-of-Speech tagging. The other modules are Term Generation and Term Filtering.

In the Term Generation stage, sequences of tagged lemmas are selected as potential term candidates on the basis of relevant morpho-syntactic patterns (such as "Noun Noun", "Noun Preposition Noun", "Adjective Noun", etc.). The candidate combination stage is performed in several passes. In each pass, association coefficient between each pair of adjacent terms is calculated and a decision is made whether they should be combined. In the case of competing possibilities (such as (t1 t2) and (t2 t3) in (t1 t2 t3)), the pair having the better association coefficient is replaced first. The documents are then updated by converting all combined terms into atomic terms by concatenating the terms with an underscore. The whole procedure is then iterated until no new terms are generated [11].

In generating terms, it is important to use a filter that preserves higher precision and recall. The corpus is tagged, and a linguistic filter will only accept specific part-of-speech sequences. The choice of the linguistic filter affects the precision and recall of the results: having a 'closed' filter, which is strict regarding the part-of-speech sequences it accepts, (like only Noun +...) will improve the precision but will have a bad effect on recall [13]. On the other hand, an 'open' filter, which accepts more part-of-speech sequences, such as prepositions, adjectives and nouns, will have the opposite result. In [14], a linguistic filter is chosen, staying somewhere in the middle, accepting strings consisting of adjectives and nouns:

*(Noun / Adjective) + Noun*

However, the choice of using this specific filter depends on the application: the construction of domain-specific dictionaries requires high coverage, and would therefore allow low precision in order to achieve high recall, while when speed is required, high quality would be better appreciated, so that the manual filtering of the extracted list of candidate terms can be as fast as possible. So, in the first case we could choose an 'open' linguistic filter (e.g. one that accepts prepositions), while in the second, a 'closed' one (e.g. one that only accepts nouns). The type of context involved in the extraction of candidate terms is also an issue. At this stage of this work, the adjectives, nouns and verbs are considered [14].

### C. Term Weighting

The Term Generation stage produces a set of terms associated with each document without taking into account the relevance of these terms in the framework of the whole document collection. The goal of term weighting is to assign to each term found a specific score that measures the importance, with respect to a certain goal, of the information represented by the term.

The experimental evidence accumulated over the past several years indicates that text indexing systems based on the assignment of appropriately weighted single terms produce retrieval results that are superior to those obtainable with other more elaborate text representations. These results depend crucially on the choice of effective term weighting systems. The main function of a term weighting system is the enhancement of retrieval effectiveness. Effective retrieval depends on two main factors: on the one hand, items likely to be relevant to the user's needs must be retrieved; on the other hand, items likely to be extraneous must be rejected [15]. The research in this area have found various term weighting schemes. In [15], the method called "TF-IDF Weighting Scheme" is described and [14] describes a method called "C-Value Method".

### III. APPROACH

The input to the proposed system is a collection of law reports. Law reports consist of two sections; namely the head and the detail section. The head section summarizes the whole law report and the detail section contains the detailed information about the case. Only the head section is used for automated processing as it contains sufficient details for the purpose.

Figure 1 shows the overall architecture of the proposed system.

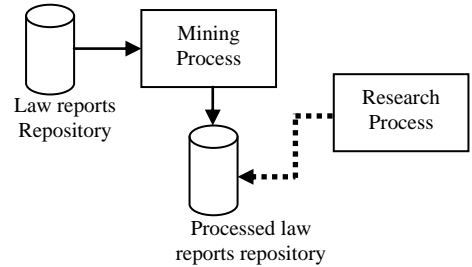

Figure 1: Overall Architecture of the Proposed System

The proposed system consists of two main components, namely;

a. The mining process
b. The research process

The mining process is the main process in the framework and to be completed prior to the research process. The mining process is carried out on the entire collection of the law reports of the repository. In this process, each document is analyzed and information that should be used for legal research is recorded in the processed law reports repository. Then the research process is carried out on the processed law reports. In this process, the text block is analyzed and the required information is extracted and compared with each law report to identify the matching reports.

The proposed approach uses text mining. More precisely, it uses terms level text mining, which is based on extracting meaningful terms from documents [11]. Each law report is represented by a set of terms characterizing the document.

### A. The Mining Process

Figure 2 shows the architecture of the mining process. The mining process goes through the stages shown in Figure 2 and is based on the approach presented in [11]. The law reports are stored in a repository and the mining process works on that. Basically, one report at a time is taken, processed and the required information are stored back along with the report.

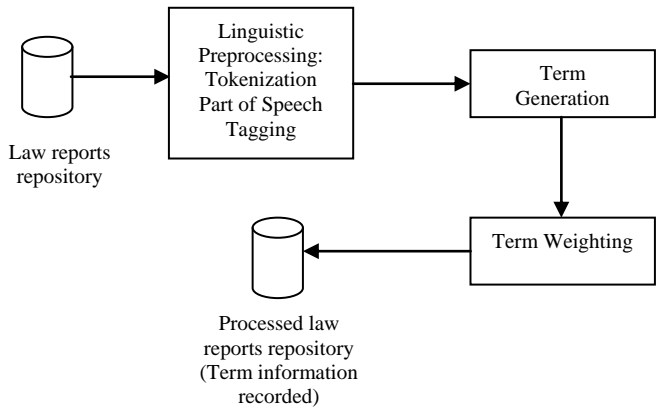

Figure 2: Architecture of the Mining Process

The first step is the Linguistic Preprocessing that includes Tokenization and Part-of-Speech tagging. Here, the head part of the law report is tokenized into parts and morpho-syntactic categories such as noun, verb, adjective etc., are associated into the words of the text. Each word in the corpus is given a grammatical tag as noun, verb, adjective, adverb or preposition, which is a prerequisite step to the next stage. A list of predefined tags is used for tagging the words. Table A-1 in Appendix lists the set of speech tags used.

Along with part-of-speech tagging, the other preprocessing stage is 'Chunking'. A chunk is a syntactically correlated part of a language (i.e. noun phrase, verb phrase, etc.). Chunking is the process of identifying those parts of language. Part-of-speech tagging and chunking are bound together. As with part-of-speech tagging, chunking also uses a tag list and is given in Table A-2 in Appendix. A chunk tag is defined in the following format.

<Prefix> - <Chunk Type>

Figure 3 shows a sample block of text that has been speech tagged and chunked.

```
Fundamental/JJ/B-NP Rights/NNS/I-NP -/:/O T
ransfer/NN/B-NP of/IN/B-PP petitioner/NN/B-
NP as/IN/B-PP Principal/NNP/B-NP ,/,/O Razi
ck/NNP/B-NP Fareed/NNP/I-NP Maha/NNP/I-NP V
idyalaya/NNP/I-NP procured/VBN/B-VP by/IN/B
-PP influence/NN/B-NP or/CC/O deceit/VB/B-V
P -/:/O Petitioner/NNP/B-NP not/RB/B-ADJP e
ligible/JJ/I-ADJP for/IN/B-PP Principal/NNP
/B-NP 's/POS/B-NP post/NN/I-NP of/IN/B-PP t
hat/DT/B-NP school/NN/I-NP -/:/O Subsequent
/JJ/B-NP appointment/NN/I-NP of/IN/B-PP eli
gible/JJ/B-NP candidate/NN/I-NP challenged/
VBD/B-VP -/:/O Article/NN/B-NP 12/CD/I-NP (
/(/I-NP 1/CD/I-NP )/)/O of/IN/B-PP the/DT/I
-NP Constitution/NNP/I-NP ././O The/DT/B-NP
 petitioner/NN/I-NP who/WP/B-NP was/VBD/B-V
```

Figure 3: A sample Part-of-Speech Tagged Text Block

## B. Term Generation

In the Term Generation stage, sequences of tagged lemmas are selected as potential term candidates on the basis of relevant morpho-syntactic patterns (such as "Noun Noun", "Adjective Noun", etc.).

For the retention of higher accuracy and recall, only two word terms are generated as the following.

(Adj | Noun) + Noun

When the term includes more number of words, then it will have higher precision but lower recall. If the term includes only one word, then it will have higher recall but lower precision. To obtain a balanced precision and recall, by staying in the middle only two word terms are used in this approach [11],[13].

A stop word is a very common word that is useless in determining the meaning of a document. Stop words do not contribute a value for the meaning when selected as a term. So, when that kind of words are included in the generated term set, those terms are removed from the term set to preserve a clean valuable term set and avoid misleading meanings. In the context of legal research, the words contained in Table 1 are considered as stop words.

TABLE I: STOP WORDS IN THE CONTEXT OF LEGAL RESEARCH

| Stop Words |
|---|
| Petitioner |
| Complainant |
| Plaintiff |
| Plaintiff-respondent |
| Court |

The Term Generation stage produces a set of terms associated with each law report without taking into account the relevance of these terms in the framework of the whole law report collection.

The goal of term weighting is to assign each term found a specific score that measures the importance, relative to a certain goal, of the information represented by the term. A term with a higher weight is a more important term then other terms. The TF-IDF weighting method presented in [11] is used in this framework due to its performance and easiness.

## C. The Research Process

Research process shown in Figure 4 comes into action when the mining process has been completed. This is the search process used in the proposed framework to find out the relevant law reports, according to the user's legal research criteria.

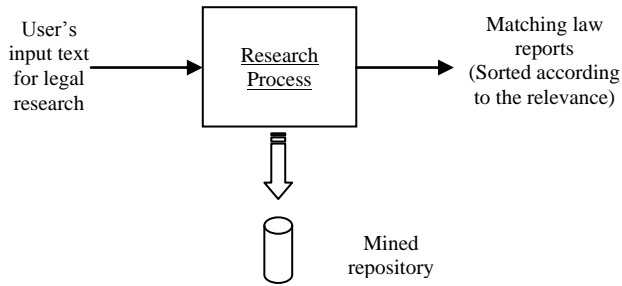

Figure 4: The Research Process

The research process is based on the input text block for the legal research given by the user. The process works on the mined repository of law reports, thus the mining process is compulsory before the research stage. The outcome is the set of matching law reports, sorted according to the relevance of the reports to the input research text.

The research process mainly works on the input text block for the legal research. A similar process as of mining the law reports is carried out on the input text block. All the steps in the mining process including linguistic pre-processing, term generation, term weighting are done on the input text. In short, the input text is treated like another document.

After generation of terms and weight assignment, the next stage is document comparison. Vector space model is used for comparing the input text (treating it as a document) against the law reports in the repository. The law reports are represented in terms of vectors (keywords) in a multi-dimensional space as shown in Figure 5, and the theory called the "Cosine Similarity" is used for comparison [16–23]. Using the cosine formula shown in Figure 6, the cosine angle between the query document and each law report in the repository is computed. Then the matching reports are sorted in descending order based on the cosine value that ranges between 0 and 1. This process brings the most relevant law report to the query, to the top.

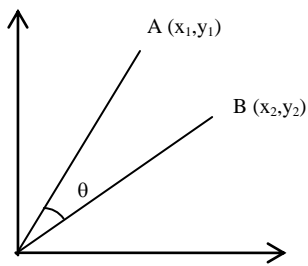

Figure 5: Representation of Documents in Vector Space

$$cos(\theta) = \frac{v_1 . v_2}{\|v_1\| \|v_2\|}$$

$$cos(\theta) = \frac{(x_1 . x_2) + (y_1 . y_2)}{(x_1^2 + y_1^2)^{1/2} + (x_2^2 + y_2^2)^{1/2}}$$

Figure 6: Formulas for Computing Cosine Similarity

## IV. PROTOTYPE IMPLEMENTATION

Prototype application was built using Java Platform, Standard Edition. For part-of-speech tagging, text processing library was used in a high-level manner, and the other parts of the architecture are implemented using object orientation.

Figure 7 shows the user interface of the software. The user interface can be divided in to two parts. The left side of the interface is used for the storage of law reports and text mining, while the right side of the interface has facilities for the research process.

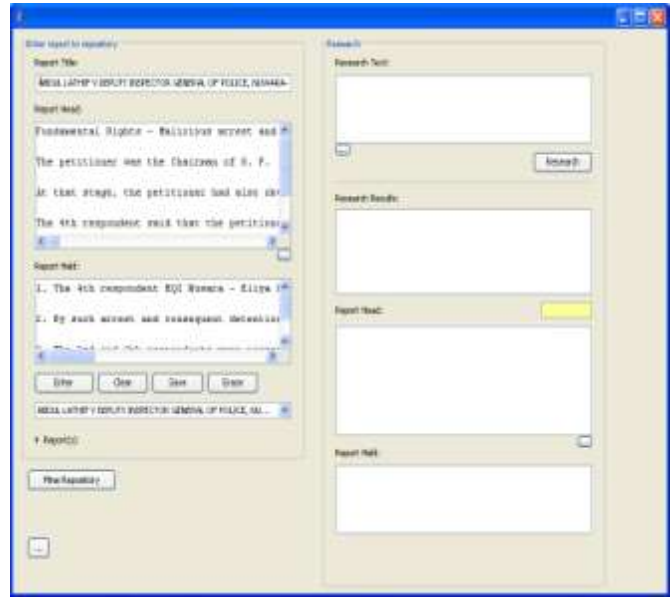

Figure 7: User Interface of the Prototype Application

The text processing library used for part-of-speech tagging uses a model based approach for the functionality. At the startup of the application, the required components are initiated and the models and other resources are loaded. Once loaded, they can be used throughout the application reducing the loading time during operations. This improves the efficiency of the operations. Figure 8 shows the resource loading at the startup.

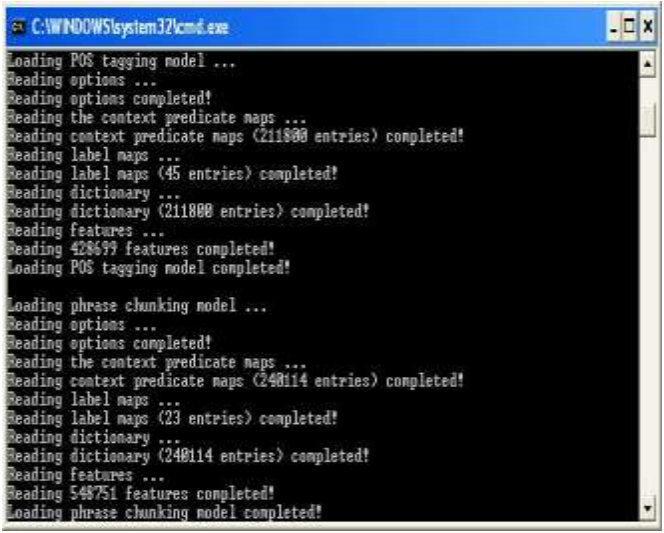

Figure 8: Resource Loading at the Startup of Application

When the law reports have been loaded to the repository, mining should be carried out. The prototype has the facilities for viewing intermediate results such as part-of-speech tagging details, terms, weights etc., However, when a new law report is loaded to the repository the mining process should be executed in order to update the mining results according to the updated state of the repository.

Figure 9 shows the result of the mining process with the intermediate results window open.

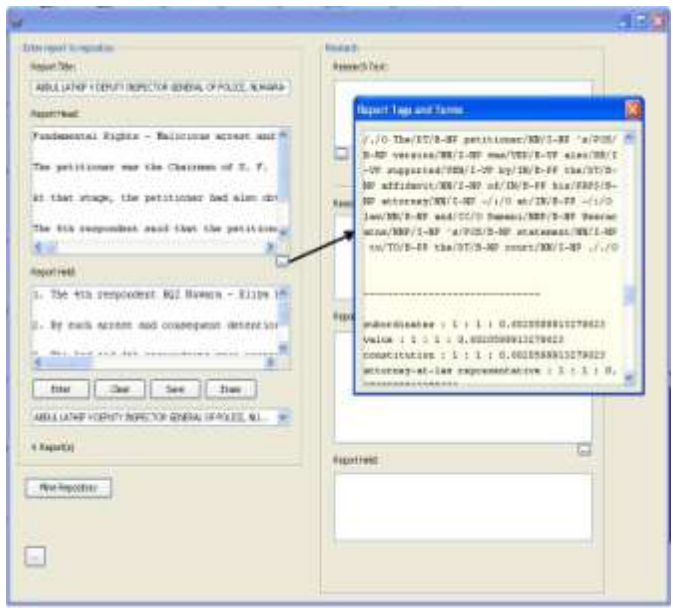

Figure 9: Results of Mining Process

Once the mining process is completed, the data is ready for the research process. The user text based on which the research process is to be carried out needs to be supplied to the application through the user interface. During the research process, first the user text will be analyzed and tagged as in the mining process. Then the law reports in the repository are compared against the user text in order to find out the relevant law reports. The interface application has the capability to show the intermediate results of the research process as well.

Figure 10 shows the results after the completion of the research process which is the last stage in the entire legal research process. Similar to the results of the mining process, the text tagging of the user text can also be viewed as intermediate results at the end of the research process. At the end of the research process, the results are presented on the same interface with all the law reports with relevant information. The law reports were organized in a descending order based on the similarity score with the most relevant law report on top. By selecting a law report in the interface, it is also possible to see the information (matching terms) based on which the user text and the law report were matched highlighted and the similarity score computed for the reports. The bottom most pane of the interface shows the verdict of the case. This makes it easier for a reader to immediately know the final judgment of the case without going through the reports separately.

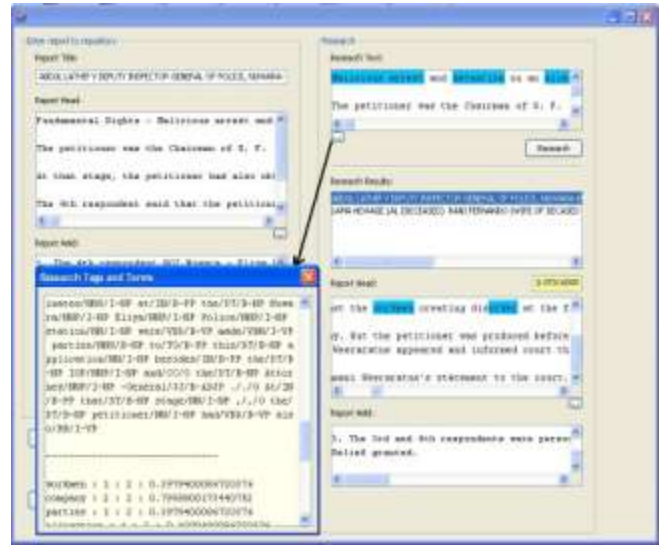

Figure 10: Final Results of the Legal Research Process

## V. EVALUATION

The evaluation of the framework was carried out using the prototype application described under Section IV. For the purpose of evaluation, only the fundamental rights cases filed at the Supreme Court of Sri Lanka were used. The reason for using the fundamental rights cases was the relative easiness of finding the case records as the Supreme Court and the Court of Appeal are superior courts and the courts of record in Sri Lanka and the Supreme Court is the final appellate court and its rulings bind all the lower courts in Sri Lanka. Also only the Supreme Court has the jurisdiction to hear fundamental rights petitions in Sri Lanka. Hence, for the fundamental rights petitions, the Supreme Court is the court of first instance.

All the other types of cases like civil cases, criminal cases and intellectual property right violation cases need to be filed at lower courts such as the District Court, High Court and the Commercial High Court respectively. These cases will be heard by the Court of Appeal and the Supreme Court as appeal applications against the verdict of lower courts. Hence, for proper legal research to be carried out, multiple case records of the same case filed and argued at different courts need to be analyzed. Hence, it was decided to concentrate only on fundamental rights cases for simplicity.

Several fundamental rights case records were downloaded in their raw forms from the Lawnet web site [24] that hosts case records of all reported cases in Sri Lanka. All these case records or law reports were input to system and the legal research process was carried out using different user input text as search text.

The results of the evaluation showed that the accuracy of the reports retrieved were high. It could also be observed that the ordering of the law reports based on the similarity score was acceptable as most of the time the most relevant case record had the highest similarity score and came on top. This shows the high precision of the system finding the correct case record based on the user input. Figure 11 shows the results of the system for precision against the manual judgment for the most relevant case record.

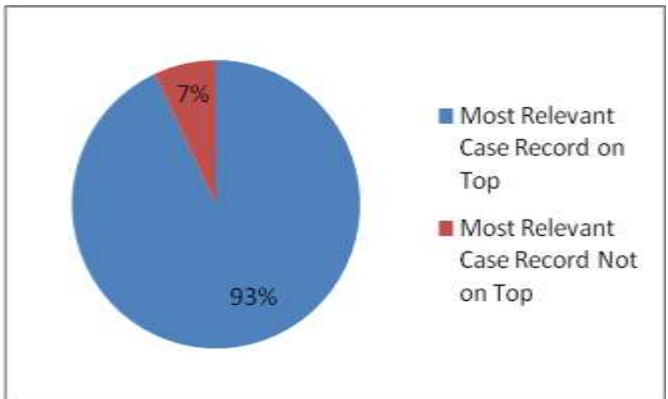

Figure 11: Result of Evaluation for Precision

When the search text was modified without changing meaning, it also resulted in the same set of case records most of the time. This shows the high recall of case records for the same or similar user input. Figure 12 shows the results of the evaluation for recall.

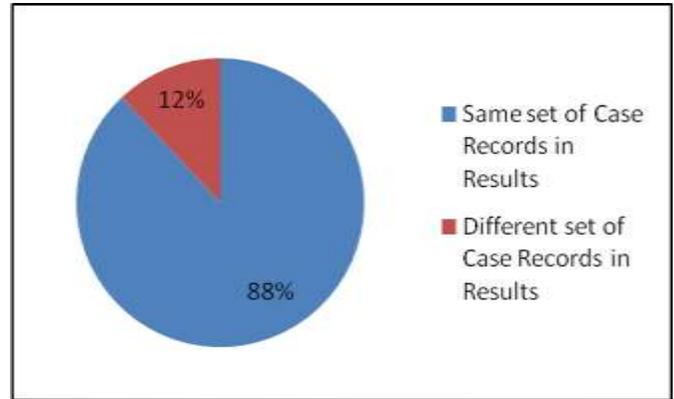

Figure 12: Results of Evaluation for Recall

## VI. CONCLUSIONS

This paper presents the results of the research carried out to develop a framework to automate the often tedious time consuming process of legal research. The end result of the research is a framework which is based on a combination of several text mining techniques. Finally the framework developed was tested for accuracy using a prototype application and fundamental rights case records. Accuracy of the results in terms of precision and recall were shown to be very high.

As the future work, this can be extended to handle all types of law reports in addition to the fundamental rights cases. The accuracy can be further enhanced by using a comprehensively updated stop words list and a set of predefined terms to be introduced along with the dropping of candidate terms. The weighting scheme can also be upgraded to include context information.

## AUTHOR PROFILE

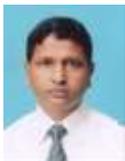
Mohamed Fazil Mohamed Firdhous is a senior lecturer attached to the Faculty of Information Technology of the University of Moratuwa, Sri Lanka. He received his BSc Eng., MSc and MBA degrees from the University of Moratuwa, Sri Lanka, Nanyang Technological University, Singapore and University of Colombo Sri Lanka respectively. In addition to his academic qualifications, he is a Chartered Engineer and a Corporate Member of the Institution of Engineers, Sri Lanka, the Institution of Engineering and Technology, United Kingdom and the International Association of Engineers. Mohamed Firdhous has several years of industry, academic and research experience in Sri Lanka, Singapore and the United States of America.

### APPENDIX

TABLE A-1: LIST OF SPEECH TAGS USED

| Tag | Description |
|-----|-------------|
| CC | Coordinating conjunction |
| CD | Cardinal number |
| DT | Determiner |
| EX | Existential *there* |
| FW | Foreign word |
| IN | Preposition or subordinating conjunction |
| JJ | Adjective |
| JJR | Adjective, comparative |
| JJS | Adjective, superlative |
| LS | List item marker |
| MD | Modal |
| NN | Noun, singular or mass |
| NNS | Noun, plural |
| NNP | Proper noun, singular |
| NNPS | Proper noun, plural |
| PDT | Predeterminer |
| POS | Possessive ending |
| PRP | Personal pronoun |
| PRP$ | Possessive pronoun |
| RB | Adverb |
| RBR | Adverb, comparative |
| RBS | Adverb, superlative |
| RP | Particle |
| SYM | Symbol |
| TO | *to* |
| UH | Interjection |
| VB | Verb, base form |
| VBD | Verb, past tense |
| VBG | Verb, gerund or present participle |
| VBN | Verb, past participle |
| VBP | Verb, non-3rd person singular present |
| VBZ | Verb, 3rd person singular present |
| WDT | Wh-determiner |
| WP | Wh-pronoun |
| WP$ | Possessive wh-pronoun |
| WRB | Wh-adverb |

TABLE A-2: LIST OF CHUNK TAGS USED

| Prefix | Chunk Type |
|--------|-----------|
| B = beginning noun phrase | NP = noun phrase |
| I = in noun phrase | VP = verb phrase |
| O = other | PP = prepositional phrase |
| | O = other |